\documentclass[10pt]{article}

\usepackage{amsmath}
\usepackage{authblk}
\usepackage[]{graphicx}
\usepackage{color}
\usepackage[latin1]{inputenc}
\usepackage{marginnote}
\usepackage[UKenglish]{babel}

\usepackage{mathbbol}
\usepackage{amssymb}             % AMS Math
\usepackage{textcomp}

\newcommand{\bK}{\mathbf{K}}
\newcommand{\bR}{\mathbf{r}}
\newcommand{\bh}{\mathbf{h}}
\newcommand{\bx}{\mathbf{x}}
\newcommand{\tA}{\tilde{A}}
\newcommand{\rmd}{{\rm{d}}}
\newcommand{\rmi}{{\rm{i}}}
\newcommand{\rme}{{\rm{e}}}

\begin{document}

\title{Defocused travelling-fringes in scanning triple-Laue x-ray interferometry}
\author[a]{C. P. Sasso}
\author[a,b]{G.Mana}
\author[a]{E.Massa}
\affil[a]{INRIM -- Istituto Nazionale di Ricerca Metrologica, strada delle cacce 91, 10135 Torino,Italy}
\affil[b]{UNITO -- Universit\`a di Torino, Dipartimento di Fisica, via Pietro Giuria 1, 10125 Torino, Italy}
%\shortauthor{Sasso, Mana, and Massa}
\maketitle

\begin{abstract}
The measurement of the silicon lattice parameter by a separate-crystal triple-Laue x-ray interferometer is a key step for the kilogram realisation by counting atoms. Since the measurement accuracy is approaching nine significant digits, a reliable model of the interferometer operation is demanded to quantify or exclude systematic errors. This paper investigates both analytically and experimentally the effect of defocus (a difference between the splitter-to-mirror distance on the one hand and the analyser-to-mirror one on the other) on the phase of the interference fringes and the measurement of the lattice parameter.
\end{abstract}

\section{Introduction}

The measurement of the silicon lattice parameter in optical wavelengths by scanning x-ray interferometry opened a broad field of metrological and science applications. In addition to realising the metre at the atomic length-scales \cite{Basile:1995}, to determining the Avogadro constant \cite{Fujii:2018}, and, nowadays, to realising the kilogram from the Planck constant, it was instrumental to the determination of the $h/m_n$ ratio \cite{Krueger:1998,Krueger:1999} and allowed the wavelength of x- and $\gamma$-rays to be referred to the metre. These links resulted in improved measurements of the deuteron binding energy and neutron mass \cite{Greene:1986,Kessler:1999} and the most accurate test of the Planck-Einstein identity $h=mc^2$ \cite{Rainville:2005}.

In 2010 and 2014, we completed measurements of the lattice parameter of a $^{28}$Si crystal used to determine the Avogadro constant and, now, to realise the kilogram by counting atoms \cite{Massa_2011,Massa_2015}. The assessment and further improvements of the measurement accuracy, approaching nine significant digits, require a reliable model of the interferometer operation to quantify or exclude parasitic contributions to the fringe phase originated by unavoidable aberrations.

The operation theory of a triple-Laue interferometer was developed in \cite{Bonse:1965,Bonse:1971,Bauspiess:1976,Bonse:1977} and refined, with particular emphasis to the aberration effects on the fringe phase, in \cite{Vittone_1994,Vittone:1997a,Vittone:1997b,Mana:2004}. In this paper, we report an experimental verification of the dynamical-theory calculation of the out-of-focus effect on the fringe phase.

The paper is organised as follows. After a short description of the experimental set-up, in section \ref{theory}, we sketch the dynamical theory of the interferometer operation. Section \ref{simulation} reports the numerical calculation of the defocus effect on the fringe phase for the interferometer cut from the $^{28}$Si crystal whose lattice parameter was an input datum for the determination of the Avogadro constant. All the computations were carried out with the aids of Mathematica \cite{Mathematica}. The relevant notebook is given as supplementary material. The measured values of the fringe-phase sensitivity to defocus are given in section \ref{test}.

\begin{figure}\centering
\includegraphics[width=\columnwidth]{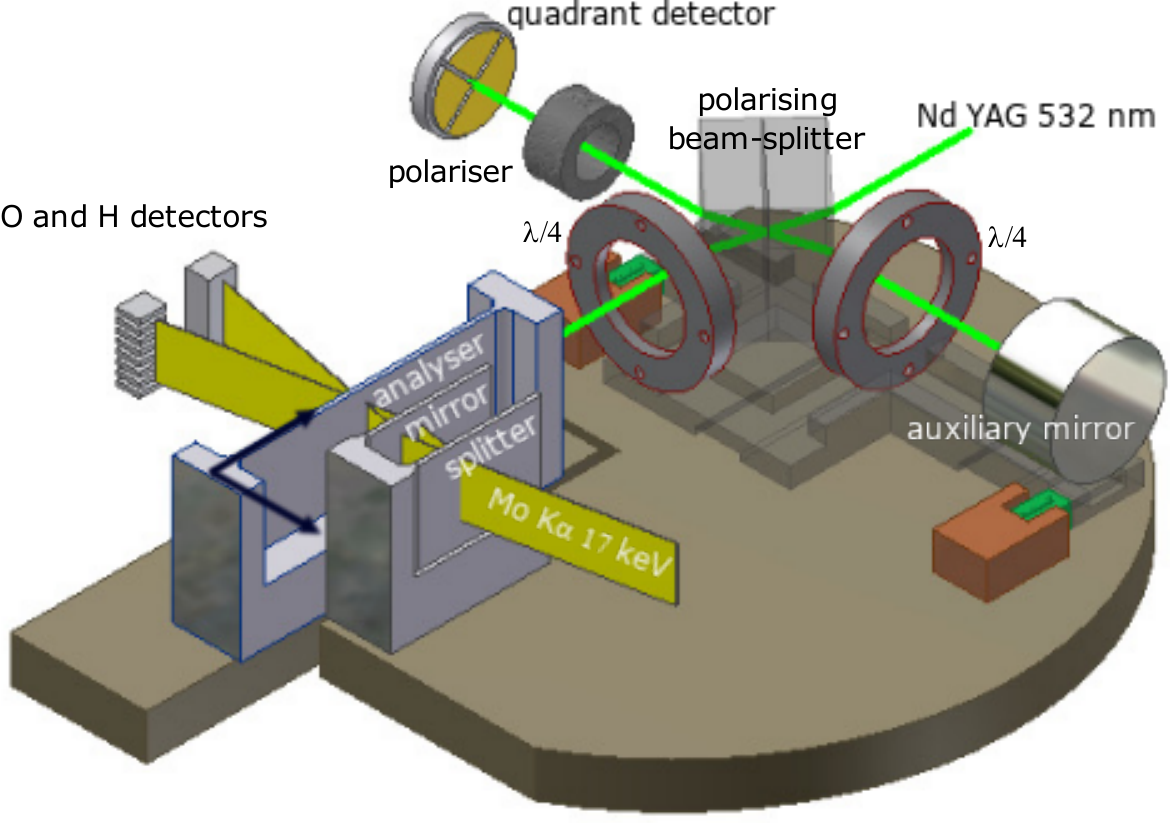}
\caption{INRIM's combined x-ray and optical in\-ter\-ferometer. The analyser displacement pitch and yaw angles are measured by laser interferometry and differential wavefront sensing. The transverse displacements (horizontal and vertical) are measured via capacitive sensors (not shown in the figure). The arrows indicate the positive directions of the axial and out-of-focus displacements.} \label{fig01}
\end{figure}

\section{X-ray interferometry}

As shown in Fig.\ \ref{fig01}, our x-ray interferometer splits and recombines, by a separate analyser crystal, Mo K$_{\alpha 1}$ x rays by Laue diffraction in perfect Si crystals. The splitter, mirror, and analyser operate in symmetric geometry, where the $\{220\}$ diffracting planes are perpendicular to the crystals' surfaces. When moving the analyser orthogonally to the diffracting planes, the interfering beams are phase shifted and travelling interference fringes are observed, the period being the plane spacing, $d_{220} \approx 192$ pm. To ensure temperature uniformity and stability and to eliminate the adverse influence of the refractive index of air, the apparatus is hosted in a (passive) thermovacuum chamber.

Detailed descriptions of the experimental apparatus are given in \cite{Bergamin:1993,Bergamin:2003,Ferroglio:2008,Massa_2011,Massa_2015,Massa:2020}. The analyser is displaced using a guide where an L-shaped carriage slides on a quasi-optical rail. An active platform with three piezoelectric legs rests on the carriage. Each leg expands vertically and shears in the two transverse directions, thus positioning the analyser over six degrees of freedom to atomic-scale accuracy. The analyser displacement, parasitic rotations (pitch, yaw, and roll), and transverse motions (horizontal and vertical) are measured via laser interferometry, differential wavefront sensing, and capacitive transducers. Feedback loops provide (axial) picometer positioning, nanoradian alignment, and axial movements with nanometer straightness.

\section{Dynamical theory of the interferometer operation}\label{theory}

The solutions of the Takagi-Taupin equations for the propagation of x-rays in perfect crystals and triple-Laue interferometers are given in \cite{Vittone:1997a,Mana:2004}. The crystal fields resembles a quantum two-level system. With a two-dimensional model, they define the Hilbert space $V_2 \otimes \mathcal{L}^2(\mathbb{R})$, where $V_2$ is a two-dimensional vector space (the space of the dispersion-surface branches) and $\mathcal{L}^2(\mathbb{R})$ is the space of the square-integrable functions.

\begin{figure}\centering
\includegraphics[width=0.9\columnwidth]{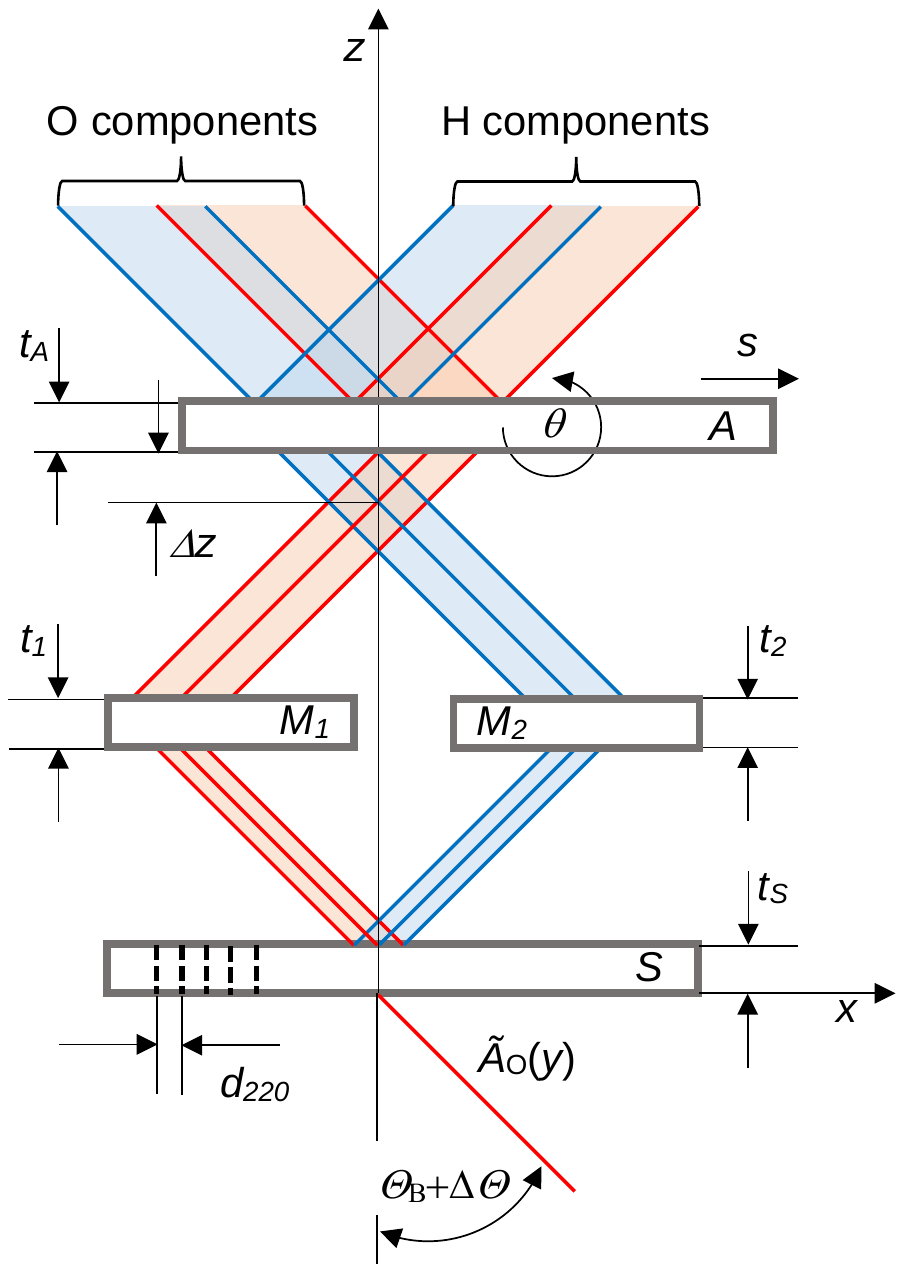}
\caption{Two-dimensional model of a symmetrical LLL interferometer. S splitter, M1 and M2 mirrors, A analyser. Red and blue rays indicated the paths 1 and 2, respectively. The $x$ axis is orthogonal to the diffracting planes. $\Theta_B$ is the Bragg angle, $\Delta z$ the defocus (positive if the analyser moves towards the negative $z$ direction), $\theta$ the analyser misalignment, $s$ the analyser displacement (positive if the analyser moves towards the positive $x$ direction).} \label{fig02}
\end{figure}

With coherent and monochromatic illumination and omission or rearrangement of common constant and phase terms, the reciprocal space representations of the waves which leave the interferometer after crossing it along paths 1 and 2 are
\begin{subequations}\begin{eqnarray}\label{1-beam}
 |D_1(y)\rangle &= &G_1 \left( \begin{array}{c}
                A_{O1}(y) \\ A_{H1}(y)
              \end{array} \right) \\ \nonumber
        & &\times \exp\left[\rmi \left( hs + \frac{(n_0-1)K\Delta t}{\cos(\Theta_B)} + \zeta y \right)\right] \\ \label{2-beam}
 |D_2(y)\rangle &= &G_2 \left( \begin{array}{c}
                A_{O2}(y) \\ A_{H2}(y)
              \end{array} \right) ,
\end{eqnarray}\end{subequations}
where $y = (\Delta_e \tan(\Theta_B)/\pi)p$ is the deviation parameter, $\zeta=2\pi\Delta z/\Delta_e$ is the dimensionless defocus, $p$ is the variable conjugate to $x$,
\begin{equation}
 G_{1,2} = \exp\left( \frac{-\mu_0 (t_s+t_{1,2}+t_A)}{2\cos(\Theta_B)} \right) ,
\end{equation}
accounts for the photoelectric absorption,
\begin{subequations}\begin{eqnarray}
 A_{O1}(y) &=& T(t_S,y) R(t_1,t) R(t_A,y+\upsilon) \tA_0(y) ,\\
 A_{O2}(y) &=& R(t_S,y) R(t_2,y) T(t_A,y+\upsilon) \tA_0(y) ,\\
 A_{H1}(y) &=& T(t_S,y) R(t_1,y) T(t_A,-y-\upsilon)\tA_0(y) ,\\
 A_{H2}(y) &=& R(t_S,y) R(t_2,y) R(t_A,y+\upsilon) \tA_0(y) ,
\end{eqnarray}\end{subequations}
are the complex amplitudes of the O and H Bloch waves $\exp(\rmi \bK_{O,H} \cdot \bR)$, $\tA_0(y)$ is the reciprocal space representation of the amplitude of the incoming Bloch's wave,
\begin{subequations}\begin{eqnarray}
 R(\tau,y) &=& \frac{\rmi \nu \sin(\tau\sqrt{y^2+\nu^2}/2)}{\sqrt{y^2+\nu^2}} , \\
 T(\tau,y) &=& \cos(\tau\sqrt{y^2+\nu^2}/2) + y R(\tau,y)/\nu ,
\end{eqnarray}\end{subequations}
are the scattering amplitudes, $\tau=2\pi t/\Delta_e$ is the di\-men\-sion\-less crystal-thickness, the indexes $\beta=\sigma, \pi$ indicating the polarisation states parallel and orthogonal to the reflection plane have been omitted. The missing symbols are given in Fig.\ \ref{fig02} and the appendix. With the convention adopted, the displacement $s$ and defocus $\Delta z$ are positive in the $x$ and $-z$ directions, respectively.

On the analyser surface, the direct- and reciprocal-space representations of the complex amplitude of the incoming Bloch's wave are
\begin{subequations}\begin{eqnarray}
 A_0(x) &\propto &\exp\left[-\frac{x^2}{2\sigma^2} + \frac{\rmi K x^2 \cos^2(\Theta_B)}{2r} \right] , \\
 \tA_0(p) &\propto &\exp \left\{ -\frac{r \sigma^2 p^2}{2[r-\rmi K\sigma^2 \cos^2(\Theta_B)]} \right\} ,
\end{eqnarray}\end{subequations}
where $K=|\bK_{O,H}|$ is the wavenumber, $\sigma$ is the beam radius and $1/r$ is the wavefront curvature.

Free propagation leads to the spatial separation of the O and H components of (\ref{1-beam}) and (\ref{2-beam}) into two localised single-component waves, which waves overlap and interfere. Detectors do not resolve the interference pattern, but measure the total photon fluxes. Consequently, an integration is necessary to describe the detected signals
\begin{equation}\label{signal}
 I_n = J_n \big[ 1 + \Gamma_n \cos(\Phi_n) \big] ,
\end{equation}
where $n=$ O, H.

Owing the limited transverse extensions of the interfering beams and large detectors, we set an infinite aperture and carry out the integration in the reciprocal space. Eventually, since photons by conventional x-ray sources have any polarisation, with equal probability, we add the $\sigma$ and $\pi$ polarisations incoherently. Therefore, in (\ref{signal}),
\begin{subequations}\begin{eqnarray}
 J_n &= &\sum_{\substack{ {\beta=\sigma,\pi} \\ {i=1,2} }}
 G_i^2 \int_{-\infty}^{+\infty} |A_{n,i}^\beta(y)|^2\, \rmd p , \\
 \Xi_n^\beta &= &G_1 G_2 \int_{-\infty}^{+\infty} A_{n,1}^\beta(y) A_{n,2}^{\beta *}(y) \rme^{\rmi y \zeta}\, \rmd p ,  \\
 \Gamma_n &= &\frac{2|\Xi_n^\sigma + \Xi_n^\pi|}{J_n} , \\
 \Phi_n &= &hs + \frac{(n_0-1)K\Delta t}{\cos(\Theta_B)} + \Psi_n \label{phase} \\
 \Psi_n &= &\arg(\Xi_n^\sigma + \Xi_n^\pi) .
\end{eqnarray}\end{subequations}

According to (\ref{phase}), the crystal displacement $s$ gives rise to travelling fringes whose period is the spacing $d_{220}=2\pi/h$ of the diffracting planes. In the reciprocal space, defocus (a difference between the splitter-to-mirror distance on the one hand and the analyser-to-mirror one on the other) shifts by $ 2\pi y \Delta z/\Delta_e$ the phase of the plane wave components travelling along the paths 1 and 2. In the real space, it shears the interfering beams by $\Delta z \tan(\Theta_B)$. With a perfect geometry (that is, $t_S=t_A$, $t_1=t_2$, and $\upsilon = \theta\Delta_e/d_{220} = 0$) and $\Delta z \ll \Delta_e$, the symmetries $\Xi_H^\beta(-\zeta)=\Xi_H^\beta(\zeta)$ and $\Xi_O^\beta(-\zeta)=-\Xi_O^\beta(\zeta)$ imply that the defocus has no effect on the phase of the H-beam fringes and changes linearly those of the O beam \cite{Vittone:1997b}.

\begin{table}
\caption{\label{parameters} Parameters used in the numerical computations. The dielectric susceptibilities are from Stepanov's x-ray server \cite{Stepanov:2004}. For the $\pi$ polarisation, we have included the $\cos(2\Theta_B)$ factor in the $h$ component of the electric susceptibility. The crystals' thicknesses are the mean values. The values in parentheses are the standard deviations of the variables assigned randomly in the Monte Carlo simulation.}
\begin{tabular}{ll}
\multicolumn{2}{l}{$\chi_0 = -3.2132 \times 10^{-6} + 1.6453 \times 10^{-8}\, \rmi $} \\
\multicolumn{2}{l}{$\chi_h^\sigma = 1.9445 \times 10^{-6} - 1.5876 \times 10^{-8}\, \rmi $ } \\
\multicolumn{2}{l}{$\chi_h^\pi = (1.8102 \times 10^{-6} - 1.4682 \times 10^{-8}\, \rmi)\cos(2\Theta_B) $} \\
$\lambda = 0.0709317$ nm    &$d_{220} = 192.014$ pm\\
$t_S = 1.196(2)$ mm         &$t_A = 1.197(2)$ mm \\
$t_1 = 1.193(2) $ mm        &$t_2 = 1.193(2)$ mm \\
$\Delta z = 0(4)$ $\mu$m    &$\theta = 0(1)$ $\mu$rad \\
$\Delta_e^\sigma = 36.29$ $\mu$m             &$\Delta_e^\pi = 41.80$ $\mu$m \\
$\mu_0 = 1.423\times 10^{-3}$ $\mu$m$^{-1}$  &$n_0-1 = 1.587\times 10^{-6}$ \\
$\kappa_\pi \approx \kappa_\sigma \approx \pi$ &$\nu_\pi \approx \nu_\sigma \approx -1$ \\
\end{tabular}
\end{table}

\section{Numerical simulation}\label{simulation}

\begin{figure}\centering
\includegraphics[width=0.9\columnwidth]{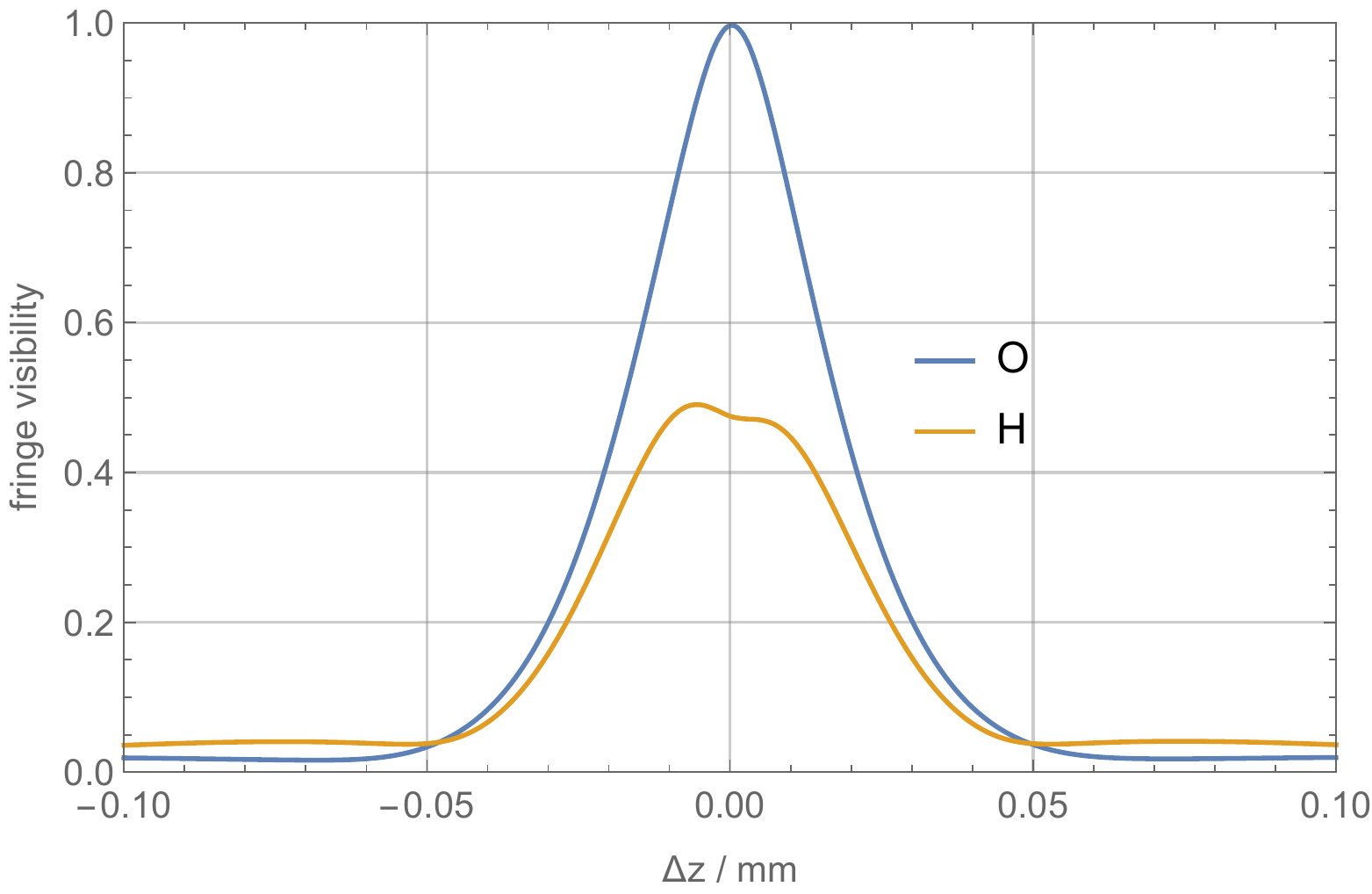}
\includegraphics[width=0.9\columnwidth]{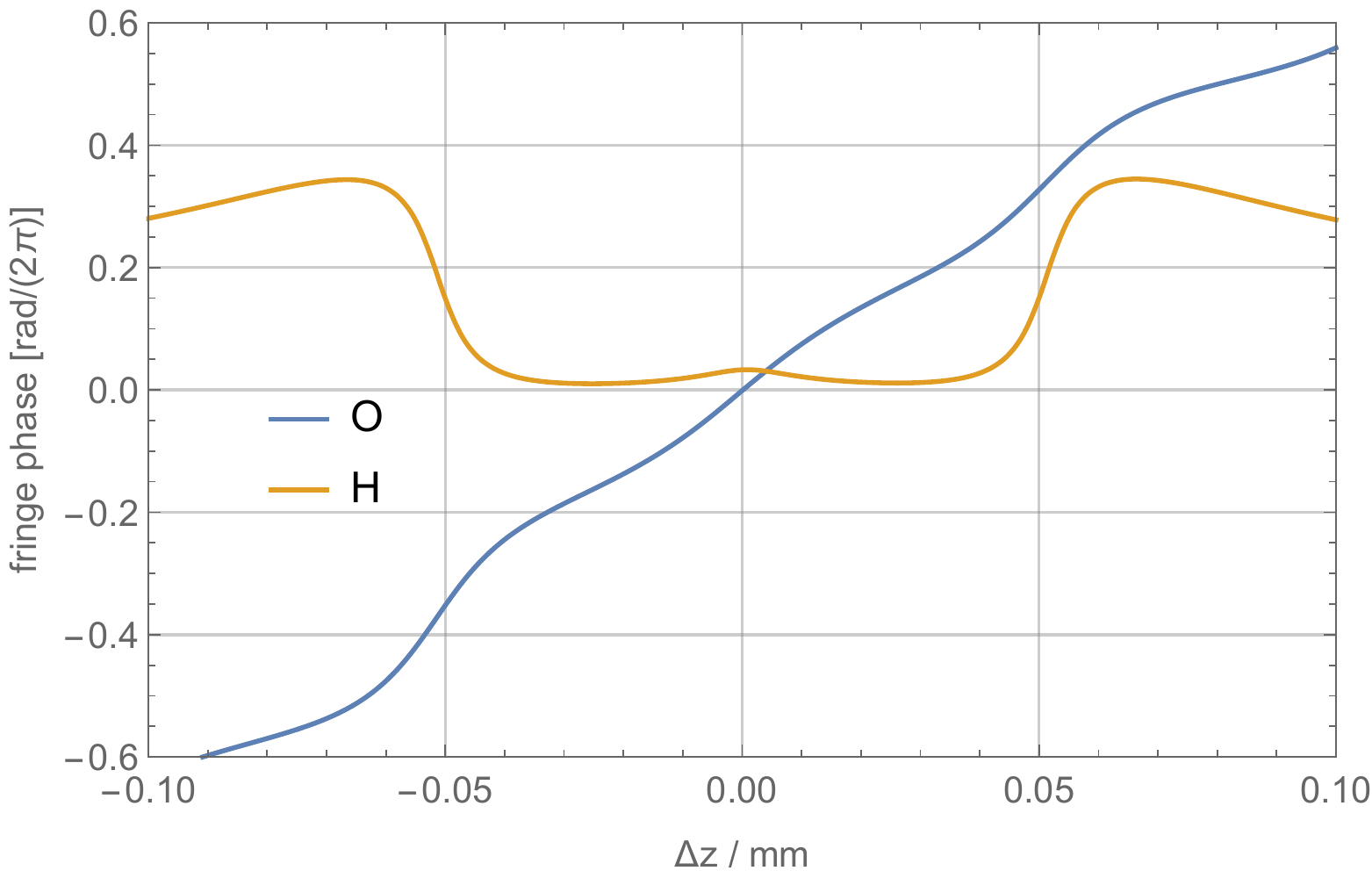}
\caption{Visibility and phase excess $\Psi_{O,H}$ of the O and H travelling fringes {\it vs.} the interferometer defocus $\Delta z$, which is positive if the analyser moves towards the negative $z$ direction, see Figs.\ \ref{fig01} and \ref{fig02}.} \label{fig03}
\end{figure}

By using the formalism developed in \cite{Vittone:1997a,Mana:2004} and outlined in section \ref{theory}, we calculated the visibility and phase of the travelling fringes as a function of the defocus. The parameters used, which refer to the interferometer used to determine the lattice parameter of $^{28}$Si \cite{Massa_2011,Massa_2015}, are listed in table \ref{parameters}. The visibility loss and phase shift are shown in Fig.\ \ref{fig03}.

\begin{figure}\centering
\includegraphics[width=0.9\columnwidth]{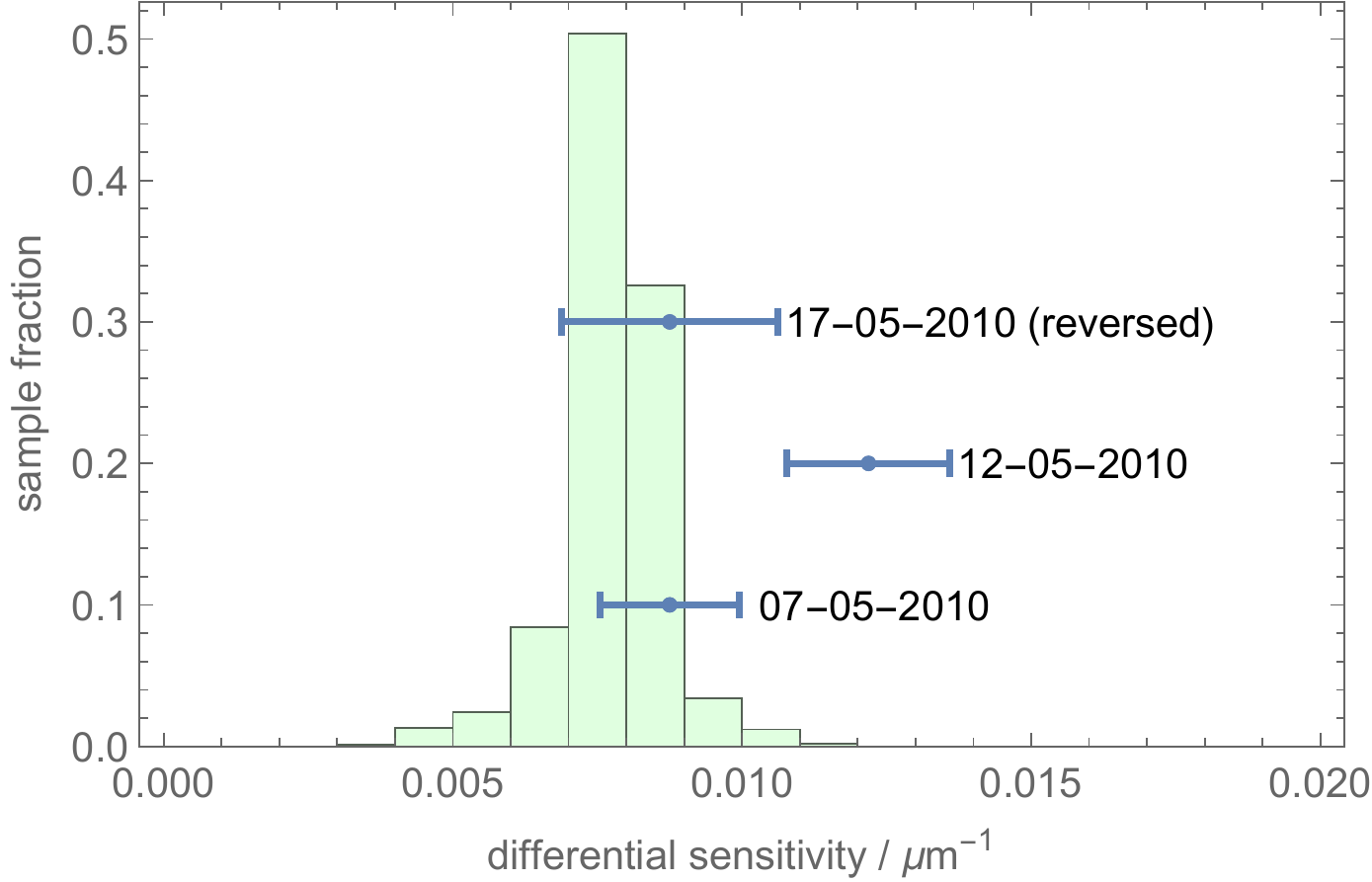}
\caption{Results of the Monte Carlo calculation the difference between the phase sensitivities of the O and H travelling fringes. The interferometer parameters and associated uncertainties given in table \ref{parameters}. Blue: observed values of the differential sensitivity $c_O-c_H$; the lines represent the standard uncertainties.} \label{fig04}
\end{figure}

The interferometer defocus contributes to the travelling-fringe phase by $2\pi c_n \Delta z$, which is valid if $\Delta z \ll \Lambda_e$, where $\Lambda_e$ is the pendell\"osung length. As shown in Fig.\ \ref{fig03}, imperfections break the visibility and phase symmetries and change the sensitivities to the defocus. To take the interferometer geometry's uncertainty into account, we evaluated the phase sensitivities to the defocus, $c_O$ and $c_H$, by a Monte Carlo simulation. Table \ref{parameters} gives the simulation parameters and the standard deviations of the normal distributions from which we repeatedly sampled the crystal thicknesses, defocus, and analyser misalignment. They have been set according to the experimental capabilities to control the interferometer geometry and alignment. The means and standard deviations of the Monte Carlo populations are $c_O = 0.0082(20)$ $\mu$m$^{-1}$ (O beam) and $c_H = 0.0004(20)$ $\mu$m$^{-1}$ (H beam).

In the next paragraph we will explain that the observable quantity is the differential sensitivity $\Delta c = c_O-c_H$, whose frequencies of occurrence in the Monte Carlo population are shown Fig.\ \ref{fig04}. The population mean and standard deviation are $\Delta c = 0.0078(9)$ $\mu$m$^{-1}$; the reduced uncertainty follows by the correlation between $c_O$ and $c_H$.

\begin{figure}\centering
\includegraphics[width=0.99\columnwidth]{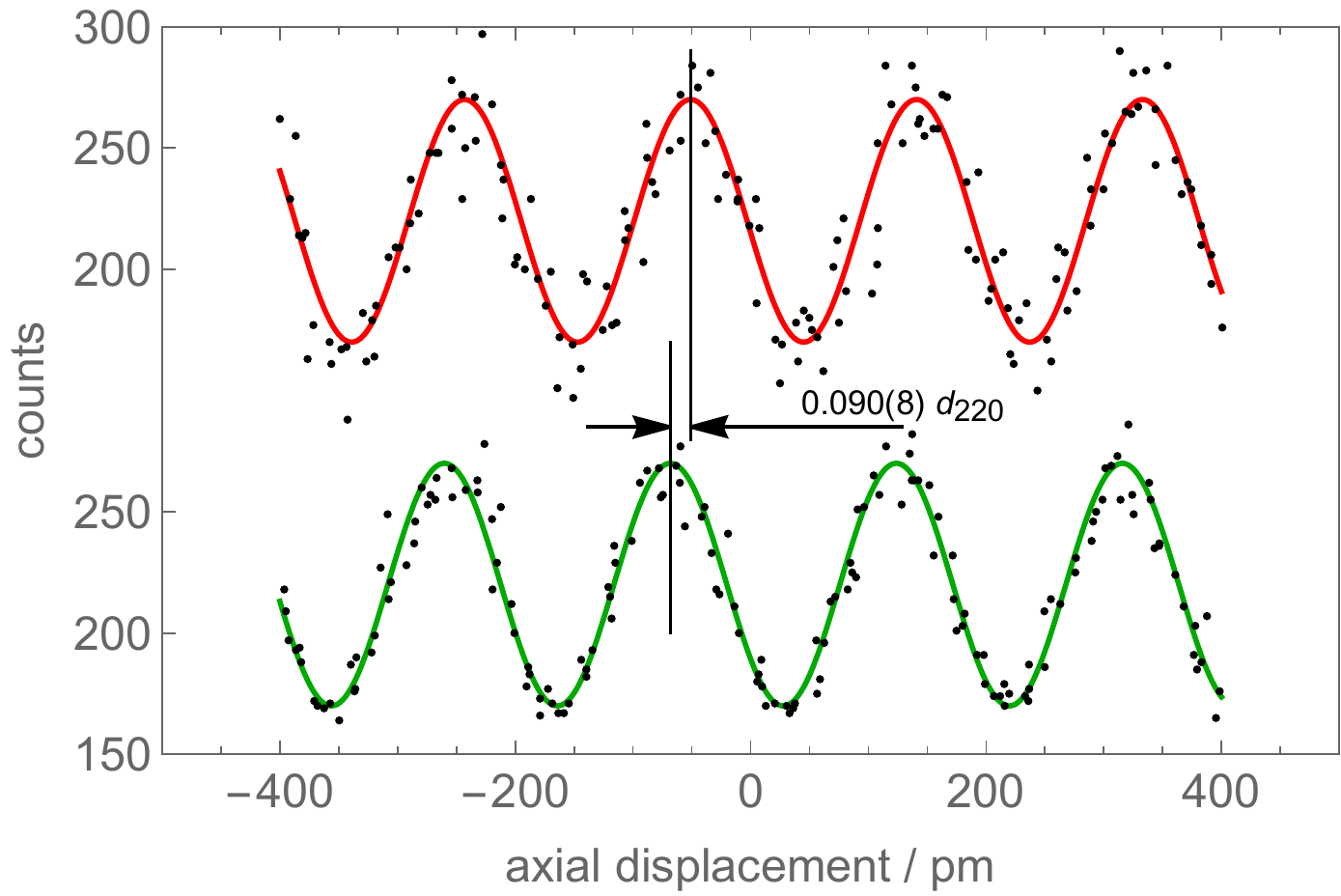}
\caption{H beam. Scans of the x-ray fringes (H beam) before (red) and after (green) a positive defocus of 3.20(15) $\mu$m (see Figs.\ \ref{fig01} and \ref{fig02}). The dots are the x-photons counted in 100 ms. The solid lines are the best-fit sinusoids approximating the data. The observed phase difference is 0.090(8) $d_{220}$ (see the first two data in Fig.\ \ref{fig06}.} \label{fig05}
\end{figure}

\section{Experimental test}\label{test}

For the experimental verification of the these predictions, we mined useful data from the archive of the lattice parameter measurements carried out in 2010. At that time, to countercheck a previous measurement of the angle between the analyser front mirror and the diffracting planes \cite{Bergamin:1999,Sasso:2021}, we defocused the interferometer by moving the analyser transversally, in a direction opposite the $z$ axis in Fig.\ \ref{fig02}, and archived the interferometer signals before and after the displacement. Because of the supporting platform's small operating range, the defocus was limited to 3.20(15) $\mu$m.

A feedback loop, relying on the laser interferometer's signals, locked to zero the axial displacement and the pitch and yaw rotations of the analyser (to within 1 pm and 1 nrad). In this way, we ensured that the translation occurred in the plane of the front mirror, which is ideally parallel to the diffracting planes.

However, a miscut angle makes the front mirror slightly misaligned and, therefore, the defocus shows a small axial component \cite{Sasso:2021}. For this reason, the only quantity experimentally observable is the difference between the phase sensitivities of the travelling fringes observed in the O and H beams. In fact, any axial displacement originates a common mode phase {colr that can be eliminated by differentiation of the phase change in the O and H branches}.

The vertical and horizontal offsets between the laser and x-ray beams were nullified to avoid differential Abbe errors. In the O beam, the interference pattern is imaged into a multianode photomultiplier through a stack of eight NaI(Tl) scintillators and identified the virtual pixel having no vertical offset. In the H beam, we imaged the whole vertical extension of the interference pattern and nullified the offset by windowing.

Since it was not possible to eliminate the drift between the optical and x-ray fringes, we implemented a modulation-demodulation strategy. Actually, we defocused repeatedly the interferometer and the two -- optical and x-ray -- signals were sampled before and after each defocusing. As shown in Fig.\ \ref{fig05}, the phases of the travelling x-ray fringes before and after defocus were recovered by least-squares estimations via the model
\begin{equation}\label{model}
 I_n(s) = J_n \big[ 1 + \Gamma_n \cos(\Omega s + \phi_n) \big] ,
\end{equation}
where $J_n$, $\Gamma_n$, $\Omega$, and $\phi_n$ are unknown parameters to be determined, $n=$ O, H, $\phi_O - \phi_H =  2\pi \Delta c \Delta z$ is the sought phase difference aiming at verifying the theoretical $\Delta c$ prediction, and the constrains $\Gamma_n >0$ and $\Omega > 0$ were applied \cite{Bergamin:1991}. The dis\-pla\-cement $s$ was positive when the analyser moves towards the positive $x$ direction, see Figs.\ \ref{fig01} and \ref{fig02}.

\begin{figure}\centering
\includegraphics[width=0.99\columnwidth]{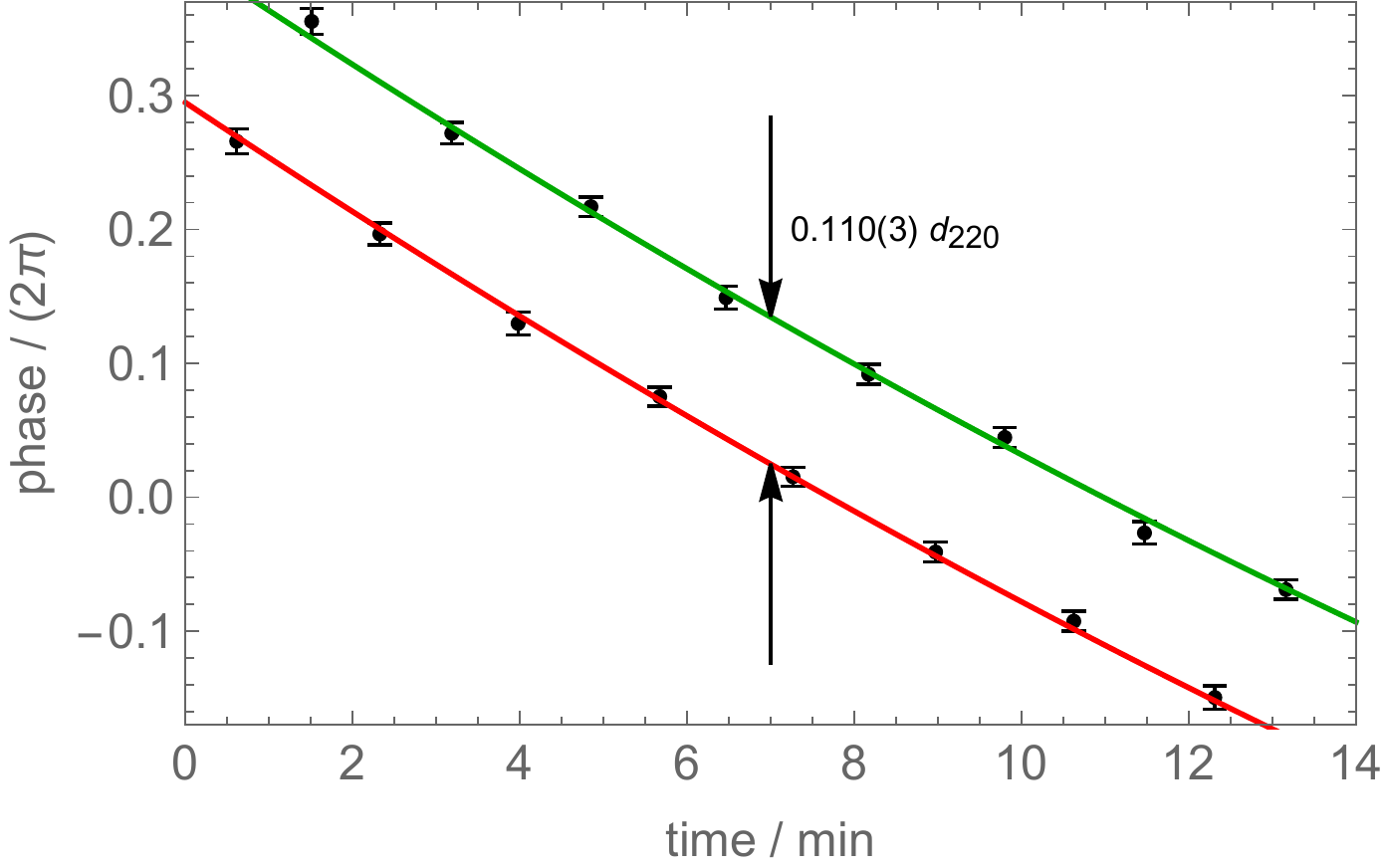}
\caption{H beam. Phases of the x-ray fringes measured before (red) and after (green) a positive defocus of 3.20(15) $\mu$m (see Figs.\ \ref{fig01} and \ref{fig02}). The dots are the measured phases (see Fig.\ \ref{fig05}). The bars are the associated standard uncertainties. The phase difference, 0.110(3) $d_{220}$, is not null because of the analyser miscut angle.} \label{fig06}
\end{figure}

\begin{figure}\centering
\includegraphics[width=0.99\columnwidth]{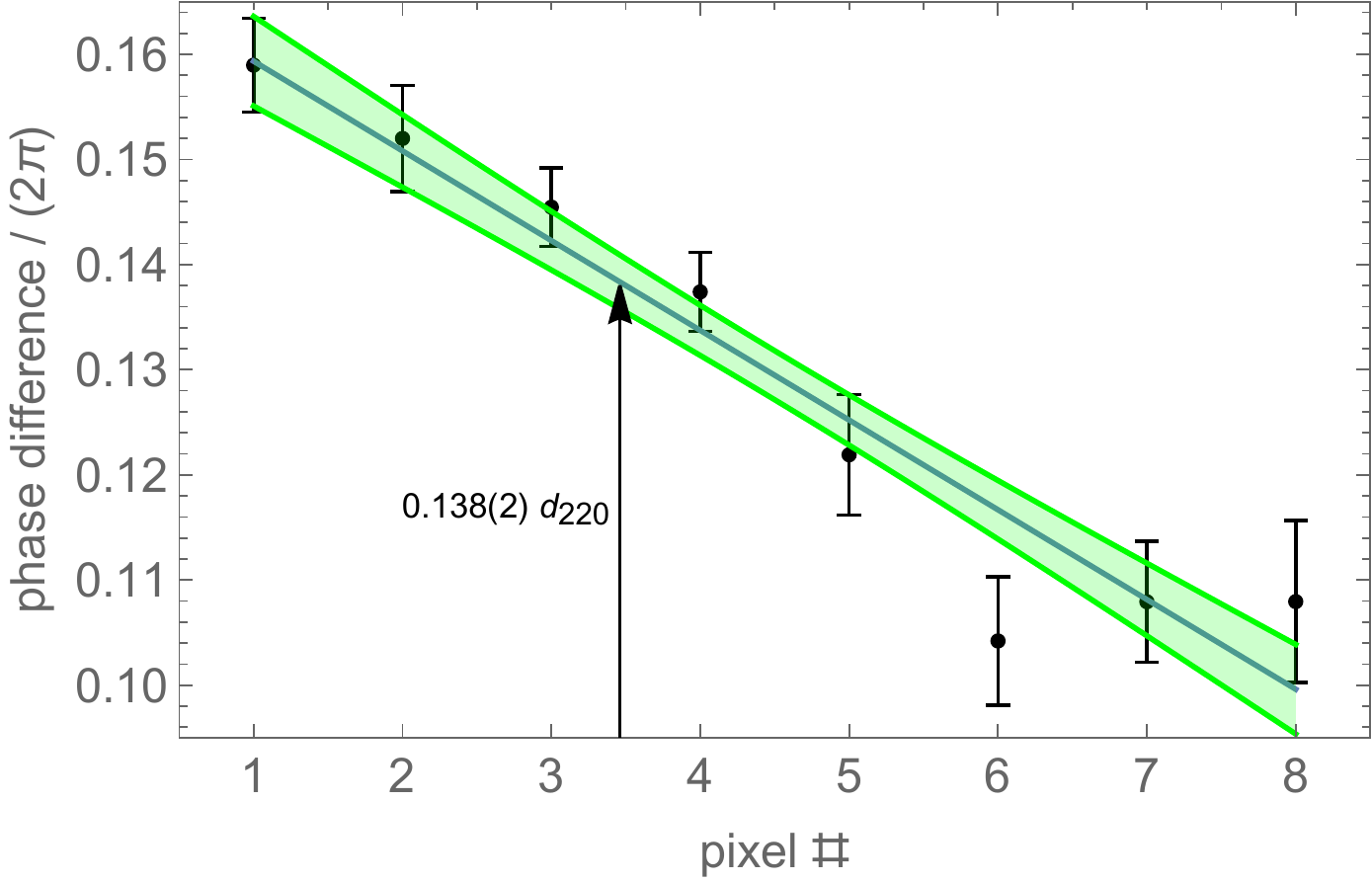}
\caption{O beam. Phase differences between the defocused -- 3.20(15) $\mu$m -- and focused travelling fringes {\it vs.} the detector pixels. The bars are the associated standard uncertainties. The solid line is the line approximating the data; the filled area indicates the standard deviation. The phase difference at the 3.46 virtual pixel having the same residual vertical-offset as the H beam is 0.138(2) $d_{220}$.} \label{fig07}
\end{figure}

Next, as shown in Fig.\ \ref{fig06}, the drift was identified and subtracted by fitting the phases of the x-ray fringes with polynomials differing only by the sought phase difference. Regarding the O beam, we calculated the phase difference between the defocused and focused fringes at the virtual pixel having the same residual vertical offset as the H beam. The result is shown in Fig.\ \ref{fig07}. The difference between the phase sensitivities to the defocus of the O and H fringes obtained from the data shown in Figs.\ \ref{fig06} and \ref{fig07} is $0.028(4)/3.20(15) \, \mu{\rm m}^{-1} = 0.0088(12) \, \mu{\rm m}^{-1}$. The phase gradient in Fig.\ \ref{fig07} is due to the second derivative of the residual angular instability of the laser interferometer, which instability is copied by the feedback loops into the analyser misalignment and whose non-linearity is not removed by the modulation-demodulation process.

To compare the predicted difference against the observed ones, we chose the positive signs of the analyser displacement and defocus in the same way in both the interferometer model (\ref{1-beam}-b) and the analysis of the experimental data (\ref{model}). The results are given in table \ref{results} and shown in Fig.\ \ref{fig04}. The measurements were carried out on May 7th, 12th, and 17th, 2010. The first two measurements were carried out at two different axial positions of the analyser, spaced by about 30 mm. We carried out the last after the analyser's reversal, which exchanged the entrance and exit surfaces.

\begin{table}
\caption{\label{results} Measured differences between the phase sensitivities to the defocus of the O and H fringes. The expected value is 0.0078(9) $\mu$m$^{-1}$.}
\begin{tabular}{ll}
date &differential sensitivity / $\mu$m$^{-1}$ \\
\hline
07-05-2010 &0.0088(12) \\
12-05-2010 &0.0122(14) \\
17-05-2010 &0.0088(19) \\
\end{tabular}
\end{table}

\section{Conclusions}

The study of the signals from a combined x-ray and optical interferometer revealed a satisfactory agreement between the observed and predicted phase shifts of the travelling x-ray fringes due to the defocus. This result is directly applicable to asses\-sing the measured values of the $^{28}$Si lattice-parameter and confirms that micrometre changes of focus were irrelevant to the error budget of our 5 cm scan \cite{Massa_2011,Massa_2015}. In fact, a mammoth parasitic defocus of 10 $\mu$m associated to a $s=5$ cm travel of the analyser will cause, in the worst O-beam case, a fractional phase error equal to $c_O d_{220}/s \approx 3 \times 10^{-10}$.

However, if aimed at a 1 nm/m fractional accuracy, measurements over sub-millimetre scans must consider the changes of focus seriously, for instance, due to an insufficient flatness of the analyser surface \cite{Birk_2020}.

Our result is also applicable to the completeness of the dynamical theory of x-ray diffraction in a perfect crystal. The unexplained discrepancy of the 2010-05-12 datum might be ascribed to an insufficient control of the interferometer operation. Besides, that time we were not aimed at testing the dynamical-theory predictions. Future experiments with larger and better-calibrated defocus are feasible and might put on a still safer footing our conclusions.

\appendix

\section{List of the symbols}

\begin{tabular}{ll}
$\bh=2\pi\hat{\bx}/d_{220}$  &reciprocal vector \\
$d_{220}$               &diffracting plane spacing \\
$s$                     &analyser displacement \\
$t_S, t_1, t_2, t_A$    &crystal thicknesses \\ &(splitter, mirrors, and analyser) \\
$\Delta t=t_1-t_2$      &differential mirror-thickness \\
$\tau=2\pi t/\Delta_e$  &dimensionless crystal thickness \\
$\Delta z$              &defocus \\
$\zeta=2\pi\Delta z/\Delta_e$  &dimensionless defocus \\
$\theta$                &analyser misalignment \\
$\bK_H = \bK_O+\bh$     &wavevectors of the O and H \\ &Bloch's waves \\
$2\bh\cdot\bK_O=h^2$    &Bragg's law \\
$\sin(\Theta_B)=-\frac{\bK_O\cdot\bh}{K_O h}$       &Bragg angle \\
$\Delta\Theta = \frac{p}{K\cos(\Theta_B)}$          &plane-wave deviation \\ &from Bragg's alignment \\
$\chi_{0,h}$            &Fourier' components of the \\ &electric susceptibilities \\
$n_0 = 1+\Re(\chi_0)/2$ &refractive index \\
$\mu_0 = \Im(\chi_0) K$ &absorption coefficient \\
$\kappa=\arg(\chi_h)$ \\
$\nu=\exp(\rmi\kappa)$ \\
$\Delta_e = \lambda\cos(\Theta_B)/|\chi_h|$             &Pendel\"osung length \\
$y=\Delta_e \tan(\Theta_B) p/\pi$  \\
$\;\;\;=\Delta\Theta \Delta_e/d_{220}$                  &deviation parameter \\
$\upsilon=\theta \Delta_e/d_{220}$                      &deviation-parameter shift\\ &(analyser misalignment)
\end{tabular}\\
\bibliographystyle{unsrt}
\bibliography{defocusing}

%\referencelist[defocusing]

\end{document}